# Kinematic Control of 2-wheeled Segway

Palak Purohit
*Department of Electrical Engineering*
*IIT Gandhinagar, India*
palak.purohit@iitgn.ac.in

Poojan Modi
*Department of Mechanical Engineering*
*IIT Gandhinagar, India*
poojan.modi@iitgn.ac.in

Udit Vyas
*Department of Electrical Engineering*
*IIT Gandhinagar, India*
udit.18110176@iitgn.ac.in

*Abstract—* The Segway is a popular self-balancing two wheeled vehicle. In this paper, we present a control mechanism for the planar Segway problem. The open loop analysis validates the fact that the system is unstable by default and there is a need to design a closed loop feedback to establish control for the system. This has been done by implementing a PD controller for the multiple-output system. Since the kinematic equations of the system are non-linear, initially the controller has been designed by linearizing the equations about the equilibrium point. Later, the response of the non-linear system is examined using the designed controller.

## I. INTRODUCTION

The Segway is a two-wheeled vehicle which is currently gaining popularity as a personal transporter amidst metropolitan roads. This makes the controller design for Segway crucial since the user's safety relies on it. In this paper, a controller has been designed for the linearized planar Segway problem. The performance of the controller has also been reviewed taking into account the non-linear dynamics of the system. The rest of the paper is divided as follows: Section II consists of the physical model, followed by the derivation of the kinematic equations and design calibrations. Section III focuses on Open Loop analysis of the system. Section IV consists of the controller design and the closed loop behaviour. Section V contains observations, conclusions, future work and limitations.

## II. PHYSICAL MODEL, ASSUMPTIONS AND CALIBRATION

### A. Segway Body

A Segway consists of a long rod, hinged to an axle, which connects two wheels. To simplify the modelling, the problem is reduced to a planar system, and therefore, the motion is considered to be unidirectional and both the wheels are treated as one. We make the following assumptions for modelling the system –

- All bodies are rigid
- The mechanical and electrical losses are zero
- There is constant contact between the wheels and the body, and they are rolling without slipping
- The angle of tilt is small, for linearizing the system (This constraint is relaxed in the further sections)
- The wheel is rolling without slipping on the surface.

$$x(t) = R\theta_w \quad (i)$$

### B. Kinematic equations

The Segway has been controlled with the help of twin motors coupled via a gear system. One of the motors applies a torque to the wheel in a forward direction and the other motor applies a proportional torque to the rod in the opposite direction so as to stabilize the rod with the proportional constant (K) decided by the gear ratios.

The dynamics of the system have been formulated with the help of D'Alembert's principle-

$$\frac{d}{dt}\left(\frac{\partial T}{\partial \dot{q}_j}\right) - \frac{\partial T}{\partial q_j} = Q_j \quad (ii)$$

The total kinetic energy of the rod was evaluated to be –

$$T = \frac{1}{2}\dot{x}^2\left[M + m + \frac{I_w}{R^2}\right] + \frac{1}{2}\dot{\theta}^2[I_r + ml^2] \quad (iii)$$

Taking $\theta_w$ & $\theta$ as the generalized coordinates, the external moments are to be found as-

$$Q_1 = T \;\&\; Q_2 = mgl\sin\theta - T \quad (iv)$$

Where $Q_1$ corresponds to $\theta_w$ and $Q_2$ corresponds to $\theta$.

Solving the equation $(ii)$ using values from $(iii)$ and $(iv)$ and the relation $(i)$, the dynamics of the Segway is as follows-

$$(k_1)\ddot{x} + (k_2\cos\theta)\ddot{\theta} = T + (k_6\sin\theta)\dot{\theta}^2$$
$$(k_3\cos\theta)\ddot{x} + (k_4)\ddot{\theta} = k_5\sin\theta - KT$$

With the following of system constants –

| $k_1$ | $k_2$ | $k_3$ | $k_4$ | $k_5$ | $k_6$ |
|---|---|---|---|---|---|
| $(M+m)R + \dfrac{I_w}{R}$ | $mlR$ | $ml$ | $I_r + ml^2$ | $mlg$ | $MlR$ |

Balancing the forces on the system in the dynamic equations, we find the equilibrium point to be $\theta = 0°$.

### C. Modes of Operation and Design Calibration

The Segway can be operated into two different modes. The first mode is where the user maintains a constant tilt angle of the rod with an intension to accelerate first and then maintain a constant velocity in forward motion. Under this mode of operation, we define the following relation between the angle of tilt, and the velocity of the Segway –

$$v(t) = 7 * \tanh(2.5 * \theta_i)$$

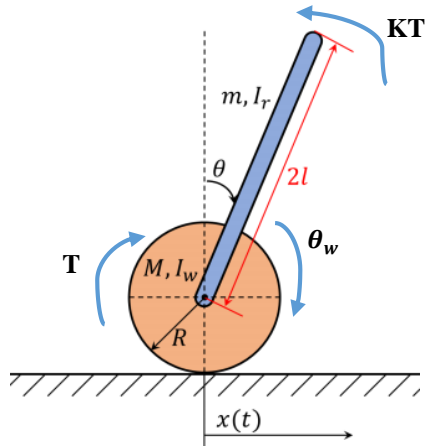

*Figure 1: Segway Planar Physical Model*

https://github.com/uditvyas/Segway_Controller.git

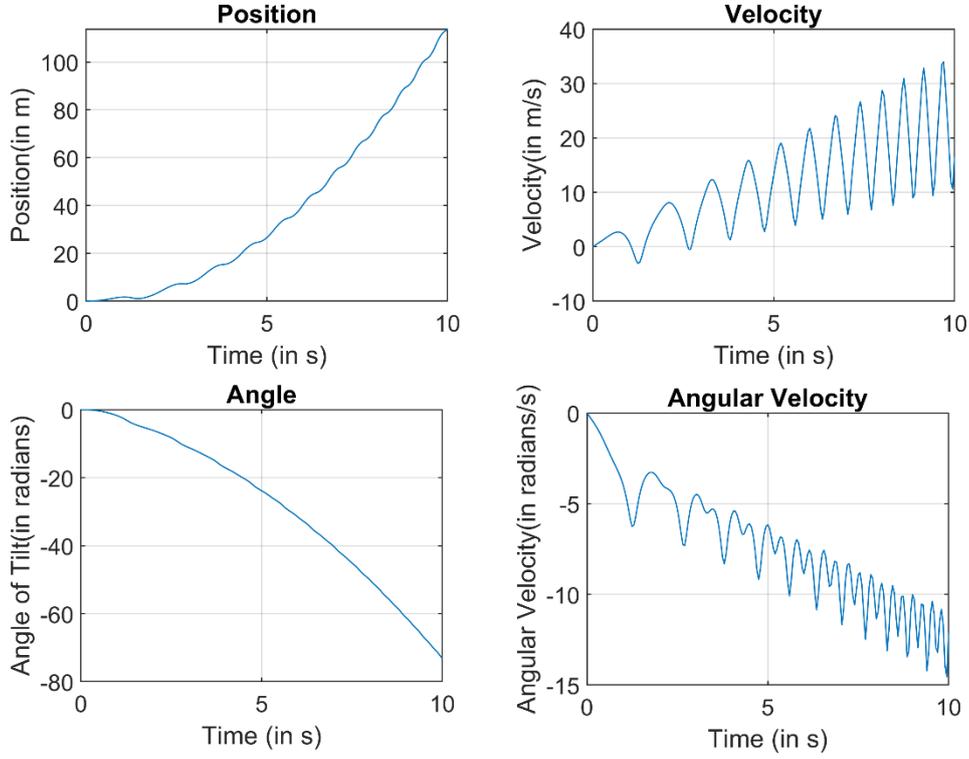

Figure 2: Non-Linear Open Loop Step Response

This step is design dependent step, and does not include any aspect of Control. The above chosen numbers are fixed based on physically intuitive values of the system states, and can be modified as per the needs of the user and feasibility. Figure 3 displays the relation between $\theta_i$ and $v_{steady}$.

The second mode of operation is when the user releases the rod, and the system is expected to come to rest. During this mode of operation, based on the velocity of the system at the instance of releasing, we estimate a stopping distance $\Delta x$ from the current position of the system. This is governed by the equation –

$$\Delta x = 3.2 * v_i$$

The linear relation is chosen keeping in mind the maximum achievable velocity, for any tilt angle. These values can be modified provided the torque requirements for the system is feasible by a suitable motor.

Our system takes in an initial tilt angle ($\theta_i$) and holding time ($t_h$) as input from the user. For the first $t_h$ seconds, the system operates in operation mode 1. After $t_h$ seconds, the system changes its operation mode to 2. It remains in this state till the system comes to rest.

### III. OPEN LOOP ANALYSIS

#### A. Non-Linear Response using ODE solver

Based on the non-linear system kinematic equations derived in Section II, we can now perform the open loop analysis on original equations, as well the linearized equations. Using the ODE solver in MATLAB, we integrate differential equations with unit step input torque to the system. This results in the plots of the system states as shown in Figure 2. The step response of the non-linear system clearly indicates that the system is unstable.

#### B. Linearization and State-Space

The given unstable system equations can be linearized about the equilibrium point ($\theta = 0°$). Thus, assuming small tilt angles, the dynamic equations derived earlier are approximated as-

$$(k_1)\ddot{x} + (k_2)\ddot{\theta} = T$$

$$(k_3)\ddot{x} + (k_4)\ddot{\theta} = k_5\theta - KT$$

Taking Laplace Transform of the above linearized equations and writing the state equations the state space for the following linear equations is thus written as –

$$\begin{bmatrix}\dot{x}\\\ddot{x}\\\dot{\theta}\\\ddot{\theta}\end{bmatrix} = \begin{bmatrix}0 & 1 & 0 & 0\\0 & 0 & -k_2k_5/(k_1k_4 - k_2k_3) & 0\\0 & 0 & 0 & 1\\0 & 0 & -k_1k_5/(k_1k_4 - k_2k_3) & 0\end{bmatrix}\begin{bmatrix}x\\\dot{x}\\\theta\\\dot{\theta}\end{bmatrix} + \begin{bmatrix}0\\(Kk_2 + k_4)/(k_1k_4 - k_2k_3)\\0\\(Kk_1 + k_3)/(k_1k_4 - k_2k_3)\end{bmatrix}T$$

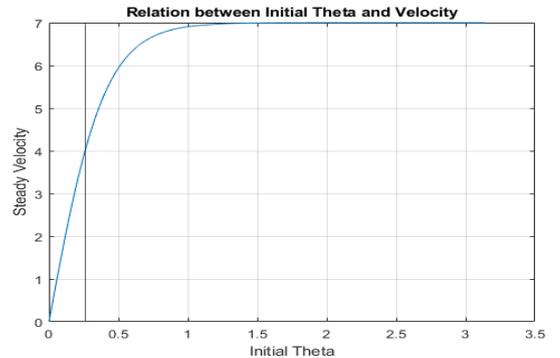

Figure 3: Relation between initial tilt angle and steady velocity

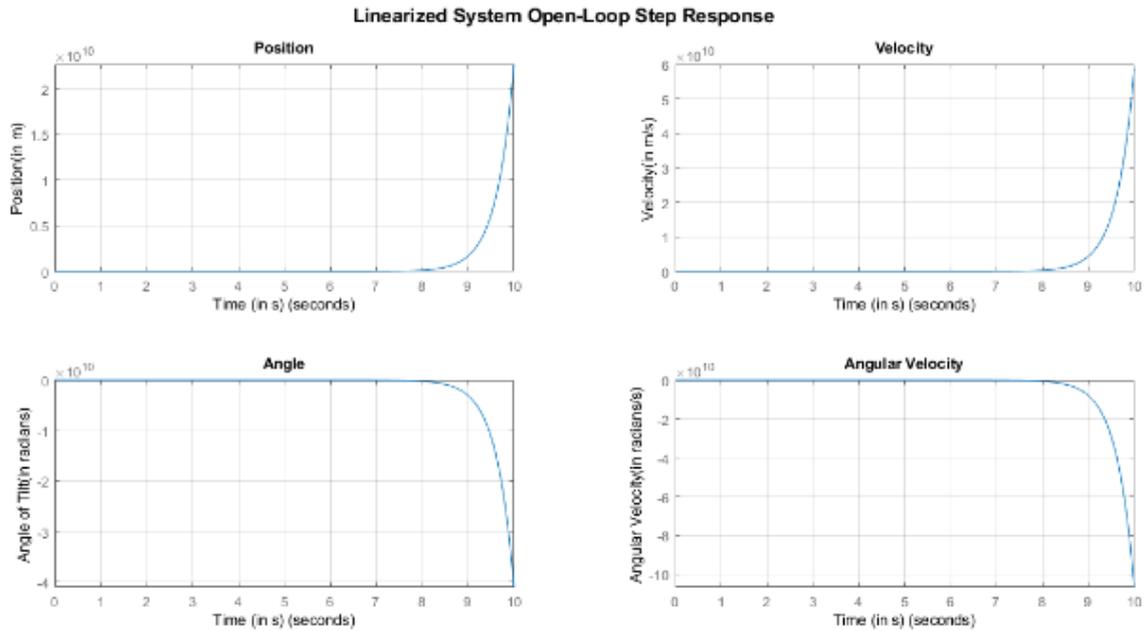

*Figure 4: Linearized Open Loop Step Response*

And the output equation is-

$$\begin{bmatrix} x \\ \dot{x} \\ \theta \\ \dot{\theta} \end{bmatrix} = \begin{bmatrix} 1 & 0 & 0 & 0 \\ 0 & 1 & 0 & 0 \\ 0 & 0 & 1 & 0 \\ 0 & 0 & 0 & 1 \end{bmatrix} \begin{bmatrix} x \\ \dot{x} \\ \theta \\ \dot{\theta} \end{bmatrix} + \begin{bmatrix} 0 \\ 0 \\ 0 \\ 0 \end{bmatrix} T$$

Thus, for the given system, we'll have four outputs namely the state variables i.e., the distance covered by the Segway wheel, the wheel velocity, the tilt angle of the rod and the angular velocity of the rod. The input for the given system will be the applied torque, T.

These four linearized relations and the subsequent transfer functions depend mainly on the value of the parameters of the system and the value of the torque proportionality constant. While the parameters are fixed, the value of the constant could be selected carefully to arrive at a suitable transfer function.

### C. Analysing Transfer Functions

In the same regard, we will have four system open loop transfer functions, each relating a system output with the system input. The corresponding transfer functions and their poles and zeros are computed and tabulated as follows –

*Table 1: Linearized transfer functions in system parameters*

| | |
|---|---|
| $G_1(s) = \dfrac{X(s)}{T(s)}$ | $\dfrac{s^2(k_4 + Kk_2) - k_5}{s^4(k_1k_4 - k_2k_3) - s^2(k_1k_5)}$ |
| $G_2(s) = \dfrac{\dot{X}(s)}{T(s)}$ | $\dfrac{s(s^2(k_4 + Kk_2) - k_5)}{s^4(k_1k_4 - k_2k_3) - s^2(k_1k_5)}$ |
| $G_3(s) = \dfrac{\theta(s)}{T(s)}$ | $\dfrac{-s^2(k_3 + Kk_1)}{s^4(k_1k_4 - k_2k_3) - s^2 k_1 k_5}$ |
| $G_4(s) = \dfrac{\dot{\theta}(s)}{T(s)}$ | $\dfrac{-s^3(k_3 + Kk_1)}{s^4(k_1k_4 - k_2k_3) - s^2 k_1 k_5}$ |

*Table 2: Poles and zeros in terms of system parameters*

| G(s) | Zeros | Poles |
|---|---|---|
| $G_1(s)$ | $\pm\sqrt{k_5/(k_4 + Kk_2)}$ | $\pm\sqrt{k_1k_5/(k_1k_4 - k_2k_3)}, 0, 0$ |
| $G_2(s)$ | $\pm\sqrt{k_5/(k_4 + Kk_2)}, 0$ | $\pm\sqrt{k_1k_5/(k_1k_4 - k_2k_3)}, 0, 0$ |
| $G_3(s)$ | 0,0 | $\pm\sqrt{k_1k_5/(k_1k_4 - k_2k_3)}$ |
| $G_4(s)$ | 0,0,0 | $\pm\sqrt{k_1k_5/(k_1k_4 - k_2k_3)},$ |

From Table 2, we can see that each of the system transfer function has a pole lying at $\sqrt{k_1k_5/(k_1k_4 - k_2k_3)}$ on the right half of the s-plane. Thus, the given system is unstable. The given pole will always remain on the right half of the plane for any given physically realizable system. Hence, the system will always be unstable irrespective of the values of various parameters.

### D. Linearised Open-Loop Step Responses

The simplified transfer functions obtained by plugging in the values of the system parameters given in Figure 1 are as follows-

*Table 3: Linearized System Transfer Functions*

| | |
|---|---|
| $G_1(s)$ | $\dfrac{4.3735\,(s - 2.062)(s + 2.062)}{s^2(s - 2.597)(s + 2.597)}$ |
| $G_2(s)$ | $\dfrac{4.3735\,s(s - 2.062)(s + 2.062)}{s^2(s - 2.597)(s + 2.597)}$ |
| $G_3(s)$ | $\dfrac{-2.927 s^2}{s^2(s - 2.597)(s + 2.597)}$ |
| $G_4(s)$ | $\dfrac{-2.927 s^3}{s^2(s - 2.597)(s + 2.597)}$ |

The poles and zeros of the above transfer functions have been tabulated as-

*Table 4: Poles and Zeros of Transfer Functions*

| $G(s)$ | $G_1(s)$ | $G_2(s)$ | $G_3(s)$ | $G_4(s)$ |
|---|---|---|---|---|
| Zeros | ±2.062 | ±2.062,0 | 0,0 | 0,0,0 |
| Poles | ±2.597,0,0 | ±2.597,0,0 | ±2.597,0,0 | ±2.597,0,0 |

While plugging the values of system parameters, the value of the proportional constant, *K* was also plugged in to be equal to 6. The reason for the same is explained further in Section IV.

The step response for the linearized transfer functions have been plotted in Figure 4. We find that the step responses of the non-linear, and the linearized system are not close to each other. However, in either case, the system is clearly unstable.

### IV. CONTROLLER DESIGN

Control for the Segway has been achieved using two controllers for the following two mode of operations:

➢ The *first mode* involves the scenario where in the user is continuously holding the rod at a constant tilt angle, in order to accelerate and reach a steady velocity. For this case, $\theta$ is at a constant value, which implies that $\dot{\theta}$ is 0. Also, as mentioned in Section II (c), this step requires the calibration of velocity as a function of $\theta$, since a user may wish to change the velocity of the Segway by changing the angle at which the rod is being held. In this mode of operation, the user is unconcerned about the position of the Segway and hence this controller has been designed essentially to regulate the velocity.
➢ The *second mode* is when the user leaves the rod and a controlled motion of the Segway is observed until it reaches a desired position. At the final equilibrium point, the velocity, $\theta$ and $\dot{\theta}$ are all zero. The final position however is calibrated as a function of the velocity of the Segway at the instant the user releases the rod for free motion, as mentioned in Section II(c). This mode thus requires all 4 states to be controlled such that they approach the desired final state in a well-balanced manner.

The assumption made here is that there exists a digital circuit which is responsible for switching between the two control systems depending on the intention of the user.

### A. Block Diagram for Establishing Control

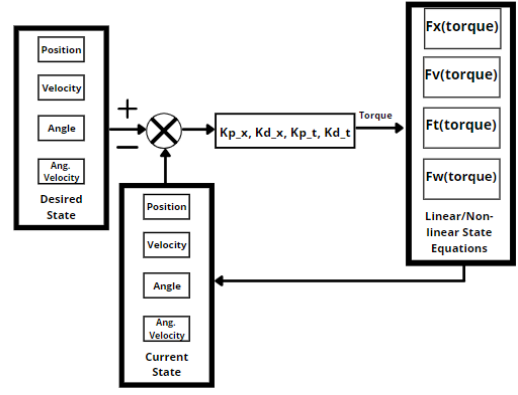

*Figure 5: Block Diagram for Control*

Figure 5 shows the basic block diagram for both controlling modes. Each state block includes position, velocity, angle and angular velocity. There exists a desired state which the Segway is supposed to achieve. The error between the current state and desired state is passed through a gain matrix which contains the following gain values:

➢ *$Kp_x$ (proportional gain for position):* The error in desired and current position is multiplied with $Kp_x$.
➢ *$Kd_x$ (derivative gain for position):* The error in desired and current velocity, i.e the derivative of position, is multiplied with $Kd_x$.
➢ *$Kp_t$ (proportional gain for angle):* The error in desired and current tilt angle is multiplied with $Kp_t$.
➢ *$Kd_t$ (derivative gain for angle):* The error in desired and current angular velocity, i.e the derivative of angular position, is multiplied with $Kd_t$.

The final value obtained after multiplication of the gain matrix and the errors is $Kp_x.e_x + Kd_x.s.e_x + Kp_t.e_t + Kd_t.s.e_t$.

To achieve the desired position, the input torque needs to be changed so as to reduce the error between the current state and the final state. Hence, this expression which takes the error for the current state into account, serves as the input torque to the system for the next state. Using the open loop transfer functions obtained in Section III(C), the torque is transformed into the position, velocity, angle and angular velocity for the next state. Again, the error is calculated between the desired state and this newly achieved state and the process goes on until the Segway reaches its desired state. On simulating, it was observed that the steady state error is negligible for all 4 states and thus the need for an integral controller was eliminated.

### B. Controllability of the System

For designing the controller, it is essential to know if the system is controllable. This can be determined by calculating the rank of the controllability matrix formed by the state space matrices given by the following state space equation.

$$\begin{bmatrix} \dot{x} \\ \ddot{x} \\ \dot{\theta} \\ \ddot{\theta} \end{bmatrix} = \begin{bmatrix} 0 & 1 & 0 & 0 \\ 0 & 0 & -k_2k_5/(k_1k_4 - k_2k_3) & 0 \\ 0 & 0 & 0 & 1 \\ 0 & 0 & -k_1k_5/(k_1k_4 - k_2k_3) & 0 \end{bmatrix} \begin{bmatrix} x \\ \dot{x} \\ \theta \\ \dot{\theta} \end{bmatrix}$$
$$+ \begin{bmatrix} 0 \\ (Kk_2 + k_4)/(k_1k_4 - k_2k_3) \\ 0 \\ (Kk_1 + k_3)/(k_1k_4 - k_2k_3) \end{bmatrix} T$$

On putting the values of the parameters, the controllability matrix formed is given below:

$$\begin{bmatrix} 0 & 4.3735 & 0 & 10.8885 \\ 4.3735 & 0 & 10.8885 & 0 \\ 0 & -2.9270 & 0 & -19.7354 \\ -2.9270 & 0 & -19.7354 & 0 \end{bmatrix}$$

Since the controllability matrix is full rank, it can be deduced that all states of the system are controllable for both modes of operation.

### C. Pole Placement for Controller Design

#### 1) Mode of Operation 1: Constant Tilt Angle

In the first mode of operation, we wish to establish control over the velocity. The open loop transfer function of velocity with respect to torque is:

$$G_2(s) = \frac{\dot{X}(s)}{T(s)} = \frac{4.3735\, s(s - 2.062)(s + 2.062)}{s^2(s - 2.597)(s + 2.597)}$$

The 4 poles have been placed as follows:

➢ The first pole has been placed at the origin, which results in pole-zero cancellation.
➢ Since the increase in velocity of the Segway should not be abrupt and jerky, the rise time should be large enough. After various parametric simulations, the two dominant poles have been chosen such that the damping ratio is 0.668 and the rise time is around 8 seconds, to provide smooth acceleration.
➢ The 4[th] pole is placed farther away at x ≈ -109.56 to substantiate the second order approximation for the dominant poles.

The characteristic equation for this pole arrangement is
$$s^4 + 110s^3 + 54.2575s^2 + 15s = 0$$

The matrix A can be represented in controllable canonical form as:
$$\begin{bmatrix} 0 & 1 & 0 & 0 \\ 0 & 0 & 1 & 0 \\ 0 & 0 & 0 & 1 \\ 0 & 0 & 6.743 & 0 \end{bmatrix}$$

This corresponds to the characteristic equation
$$s^4 - 6.743s^2 = 0$$

Comparing the coefficients for the desired characteristic polynomial and the actual one, we get:

$K_{canon} = [0, 15, 61, 110]$ which are the 4 gain values for the canonical form. The value of gains for the actual states can be found out by applying the transformation

$$K = K_{canon} * Cx * Cz^{-1}$$

Here, $C_x$ and $C_z$ represent the controllability matrices for the canonical form and the original form, respectively. The value of K comes out to be $[0, -0.8064, -21.5634, -38.7861]$.

As intended, this gives $Kp\_x = 0$, implying that there is no control on position for this mode of operation and the input torque for each step is independent of error in desired and current position. For this mode, the desired state is characterized by the final velocity which is calculated as described in Section I(C) depending on the constant tilt angle. The error in $\theta$ and $\dot{\theta}$ for each step is zero since they are maintained at a constant desired value.

#### 2) Mode of Operation 2: Free Motion of Segway

In this mode, all 4 states need to achieve a desired final value. The final equilibrium position will depend on the velocity of the Segway at the instant at which the user leaves the rod and the 2[nd] mode of operation begins, as stated in Section I(C), while the final velocity, tilt angle and angular velocity will be zero. Hence, the error in all 4 states needs to be considered while calculating the value of input torque for the next step.

Since the calculation of gains for governing a multiple output system is not straightforward as in Mode 1, the gain matrix for this mode was formed by fine tuning the gain matrix obtained in Mode 1 so as to achieve a desirable performance in terms of settling time and overshoot for all the 4 states. Extensive simulations were carried out and by analyzing the plots of position, velocity, tilt angle and angular velocity during the free motion, the following changes were made in the gain matrix calculated for mode 1:

➢ To decrease the steady state error in final position, $Kp\_x$ is increased. This leads to an increase in overshoot. To compensate for this increase and to decrease the settling time in order to reach the equilibrium position faster, the value of $Kd\_x$ is also increased. Changing these two values ensures that the position and velocity reach their desired state in a smooth and sufficiently quick manner.
➢ Since the torque experienced on the rod is proportional to the torque on the wheel but is opposite in direction, the variation of tilt angle and velocity have a close correlation. If the penalization on error of $\theta$ is increased, the system tries to decrease this error first, which leads to an increase in the time in which the velocity settles at 0 and vice versa. To account for this, the values of the proportionality constant of the torque (K) on the rod and $Kp\_t$ and $Kv\_t$ have been adjusted to achieve a reasonable settling time for both $\theta$ and velocity. This trade-off, however, leads to a minor overshoot in the tilt angle towards the negative direction. Since the value of this overshoot is insignificant and does not hamper the performance of the controller majorly, it can be neglected.

| Parameter | Rise-time | Overshoot | Settling time | Steady State Error |
|---|---|---|---|---|
| $K_p$ | Decrease | Increase | Small Change | Decrease |
| $K_i$ | Decrease | Increase | Increase | Eliminate |
| $K_d$ | Minor changes | Decrease | Decrease | No effect |

*Figure 6: Variation of response with parameters*

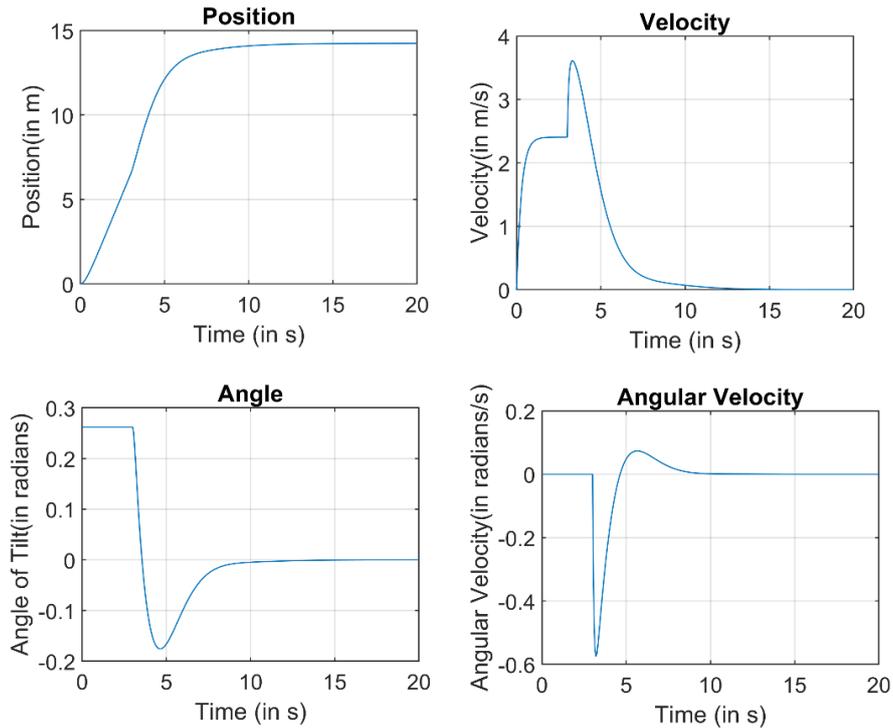

*Figure 7: Closed Loop Linear System Response for initial angle pi/12*

Based on these observations, the gain matrix comes out to be $K_{canon}$= [9, 30, 38, 15] and the corresponding K matrix for the relevant states is [-0.4839, -1.6129, -13.7056, -7.5347].

### D. Simulations for Closed Loop Linear System

The results of simulations performed for the closed loop linear system for an initial angle of $\pi/12$, if the Segway is operated in Mode 1 for 3 seconds and Mode 2 for the remaining time, has been shown in Figure 7.
The following are the inferences from the graphs:
- Position v/s Time: The position increases almost linearly for the first 3 seconds till about 7 meters when the Segway is in Mode 1. This is desired since the rod is at a constant angle which implies that the user wants to keep moving forward. It settles at t=12 seconds, that is 9 seconds after the rod has been left and stops at around 14 meters. This response is appropriate since both the settling time and the final position are suitable from the point of view of the user.
- Velocity v/s Time: From the plot it is evident that as intended, for the first 3 seconds, the Segway initially accelerates and then reaches a constant velocity of around 2.5m/s, as the rod is held at a certain tilt angle. When the user lets go of the rod, a slight jerk is experienced for about 1 second after which the velocity gradually settles down to 0 at around t=12 seconds. The jerk can be attributed to the sudden torque experienced just as the user leaves the rod, in order to reduce the tilt angle. However, the jerk is insignificant and for a very small duration. Hence, the velocity response is also apt in terms of user experience.

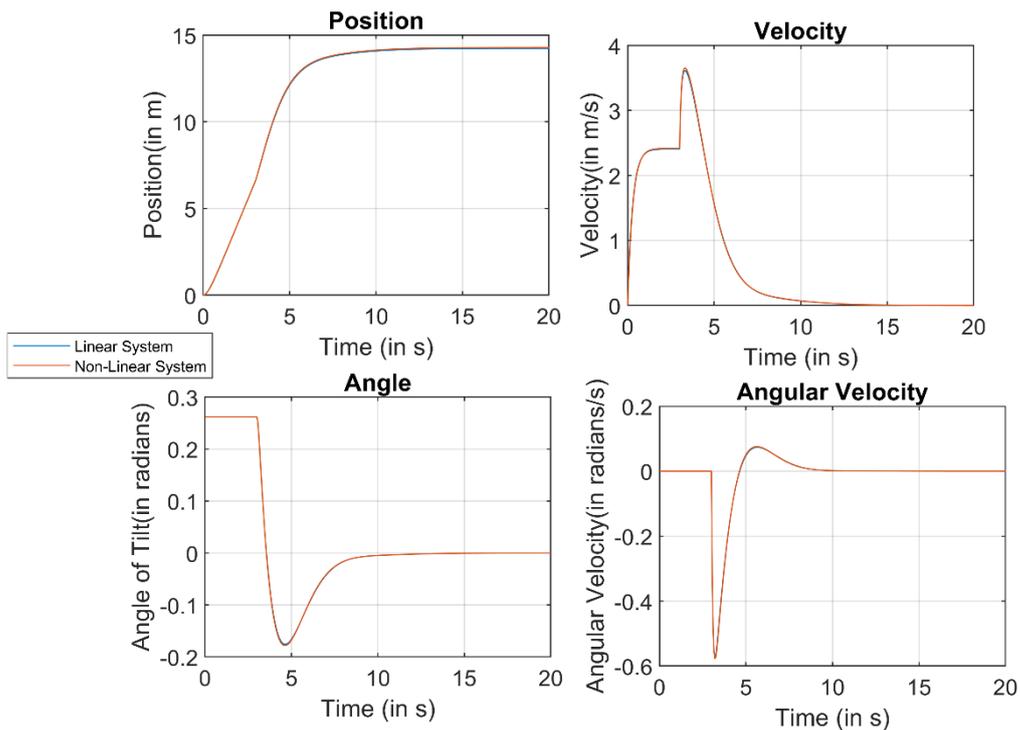

*Figure 8: Closed Loop Non-Linear System Response for initial angle = pi/12*

- Tilt Angle v/s Time: Since the tilt angle is kept at a constant value for the first 3 seconds, the graph is a straight line. When mode 2 begins, the angle starts reducing. However, for the reasons explained in Section IV C-2, the trade-off between lesser settling time for velocity and tilt angle causes a slight overshoot of the angle in the negative direction. However, this is necessary for the system to reach its equilibrium state sooner. The angle finally settles at 0 at around t= 9 seconds.
- Angular Velocity v/s Time: The angular velocity is initially 0 for the first 3 seconds since the angle is at a constant value. After the Segway enters mode 2, the value of tilt angle immediately decreases, leading to a spike in angular velocity in the negative direction. As the tilt angle comes back to 0, a positive angular velocity is seen for some time before it settles at 0.

### E. Simulations for Closed Loop Non-Linear System

The controller designed for linearized equations has been used to simulate the non-linear dynamics as well. From Figure 8, it can be made out that the difference between linear and non-linear responses is almost negligible for an initial angle of $\pi/12$ since the small angle approximation holds true. However, even for larger angles like $\pi/6$ (Fig9), the variation is almost the same. This implies that the same controller can also be used for the realistic non-linear Segway problem.

### V. CONCLUSION, LIMITATIONS & FUTURE WORK

#### A. Conclusions

The following study was able to successfully achieve closed-loop control for a Segway with the help of a feedback controller designed via state-space and pole placement methods.

- The controller was able to reach to a desired position via a more realistically viable two mode control wherein a desired velocity achieved in the first mode was then calibrated to reach a desired position.

- While holding the Segway at an initial tilt angle of 15°, the Segway was able to achieve a achieve a steady velocity of roughly 2.5 m/s after roughly 2 seconds which is in correspondence with the 1st mode of operation. Upon releasing the Segway at 3 seconds, the Segway was able to reach a final position of roughly 15 meters where it came to rest after 12 seconds. This response in our correspondence to the 2nd mode of operation. The Segway was able to achieve this control while adhering to appropriate settling and rise time values.

- While the controller was designed for linearized equations, it was able to achieve immensely similar system response for non-linear equations as well for significantly high values of theta.

- The controller was also able to minimize the jerk and discontinuity between the two modes to an extent and ensure that it is not substantial so that the rider experiences no major discomfort.

The given controller design while taking inspiration from some earlier research studies in terms of dynamics, has aimed at designing a unique controller with certain limitations and further scopes of improvement.

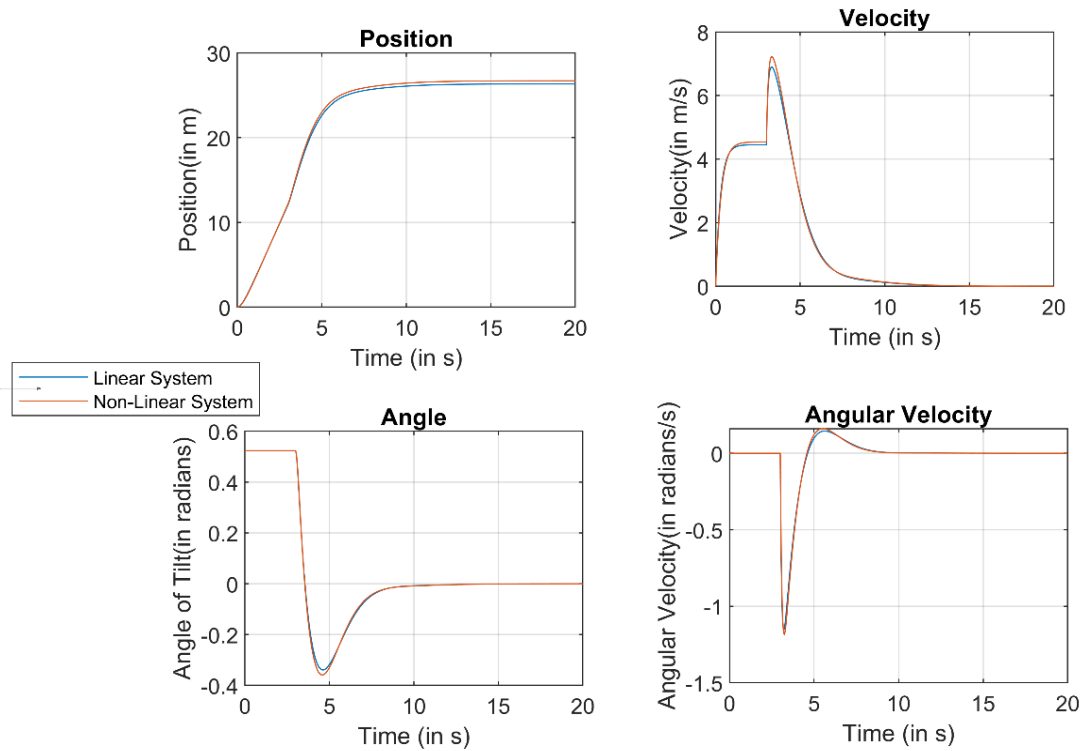

*Figure 9: Closed Loop Non-Linear System response for initial angle = pi/6*

*B. Limitations*

As concluded in the previous section, while the given controller design is a feasible one which provides a reasonable performance, there are certain limitations that need to be addressed. These are as follows-

- The given system is singly actuated. Thus, there is an underlying trade-off in terms of achieving control over position and velocity as compared to tilt angle and angular velocity of the rod.
- A physical realization of the Segway model has not been made.
- There is a slight jerk which the rider will experience while switching from mode 1 to 2. This jerk will be seen in both the velocity as well as the tilt velocity.
- As mentioned in the previous section, the controller starts to deviate from the linear assumptions for higher values of initial tilt angles. The Segway goes slightly farther and the jerk is slightly higher for values of initial tilt as high as 30°.
- Apart from the system parameters, the torque proportionality constant also depends on the weight of the rider and the frictional forces, the effect of which have not been included in the study.
- Just like a fine-tuned design calibration made in terms of Segway velocity for the 1$^{st}$ mode of operation, the calibration function for the second mode can be improved.
- The given study doesn't incorporate the lag included in reaching to the initial tilt angle.

*C. Future Outlook*

The given study is aimed at designing and analysing a two-mode controller for a Segway. The following further developments could be looked upon in the following study-

- A physical realization of the model could be made which inculcates the effects endured by the user.
- A doubly actuated system could be used to eliminate the said trade-off between positional and angular terms.
- A more comprehensive dynamic model which includes frictional losses and the work done by the centripetal forces could be developed.
- A more fine-tuned design calibration could be made for the 2$^{nd}$ mode instead of a linear relationship.
- A limiting force based on an extreme level of tilt angle could be made to avoid toppling.

ACKNOWLEDGMENT

This project has been done under the guidance of Prof. Madhu Vadali, Assistant Professor, Electrical and Mechanical Engineering, IIT Gandhinagar. We would like to extend our gratitude to him for giving us the opportunity to work on this project which has greatly enhanced our learning in the field of Control Theory in Engineering.